\documentclass[12pt,preprint]{aastex}

\usepackage{caption}

\def\bibfiles{biblio}
\def\aareferences{\bibliographystyle{apj}
                  \bibliography{aajour,\bibfiles}}

\def\rmit#1{{\it #1}}              

\def\ie{\rmit{i.e.}}

\usepackage{natbib,graphicx}

\usepackage[lofdepth,lotdepth]{subfig}
\usepackage{color}

\shorttitle{Fast-to-Alfv\'en conversion in sunspots}

\shortauthors{Felipe et al.}

\begin{document}

\title{Three-dimensional numerical simulations of fast-to-Alfv\'en conversion in sunspots}

\author{T. Felipe\altaffilmark{1}}
\email{tobias@cora.nwra.com}

\altaffiltext{1}{NorthWest Research Associates, Colorado Research Associates, Boulder, CO 80301, USA}

\begin{abstract}
The conversion of fast waves to the Alfv\'en mode in a realistic sunspot atmosphere is studied through three-dimensional numerical simulations. An upward propagating fast acoustic wave is excited in the high-$\beta$ region of the model. The new wave modes generated at the conversion layer are analyzed from the projections of the velocity and magnetic field in their characteristic directions, and the computation of their wave energy and fluxes. The analysis reveals that the maximum efficiency of the conversion to the slow mode is obtained for inclinations of 25 degrees and low azimuths, while the Alfv\'en wave conversions peaks at high inclinations and azimuths between 50 and 120 degrees. Downward propagating Alfv\'en waves appear at the regions of the sunspot where the orientation of the magnetic field is in the direction opposite to the wave propagation, since at these locations the Alfv\'en wave couples better with the downgoing fast magnetic wave which are reflected due to the gradients of the Alfv\'en speed. The simulations shows that the Alfv\'en energy at the chromosphere is comparable to the acoustic energy of the slow mode, being even higher at high inclined magnetic fields.
\end{abstract}



\section{Introduction}
\label{sect:introduction}

The study of solar oscillations has proven to be a powerful tool to infer the properties of the solar interior.  Global helioseismology, based on the interpretation of the eigenfrequencies of the resonant modes of oscillations, has provided a robust description of the internal structure and dynamics of the Sun. In the last years, these studies have been complemented by focusing on local features. Local helioseismology interprets the full wave field observed at the surface. Several complemetary techniques have been developed to probe local perturbations, as Fourier-Hankel spectral analysis \citep{Braun1995}, ring-diagram analysis \citep{Hill1988}, time-distance helioseismology \citep{Duvall+etal1993}, and acoustic holography \citep{Lindsey+Braun1990}. 

For an accurate interpretation of the measurements obtained from all these procedures, it is extremely important to achieve a deep understanding of the physics involved in the wave propagation. Due to the increasing interest of helioseismologists in active regions, such as sunspots, the knowledge of the inteaction of waves with magnetic structures has been greatly developed in the recent years. The conversion from fast-mode high-$\beta$ acoustic waves to slow-mode low-$\beta$ waves in solar active region is well-understood from the theoretical point of view \citep{Cally+Bogdan1993, Crouch+Cally2003, Schunker+Cally2006, Cally2006}. An example of the success of this analytical development is the comparison of the observational absorption and phase shift data \citep{Braun+etal1988, Braun1995} with the modeled results obtained from the conversion \citep{Cally+etal2003, Crouch+etal2005}. Numerically, the fast-to-slow conversion is also well studied in sunspot-like atmospheres \citep{Bogdan+etal2003, Rosenthal+etal2002, Khomenko+Collados2006, Khomenko+etal2009, Felipe+etal2010a}, as well as flux tubes \citep{Hasan+etal2003, Hasan+Ulmschneider2004}. See \citet{Khomenko2009} for a review.

Not all the developments of phenomena associated with wave propagation in magnetic structures have reached the same degree of maturity. An insufficiently modelled effect is the fast-to-Alfv\'en conversion. Pure Alfv\'en waves can only exist in those mediums which are homogeneous in the direction perpendicular to both magnetic field and wavenumber. In general, this will not be the situation in a gravitationally stratifed atmosphere with complex magnetic field structure, although we will simply refer to Alfv\'en waves even in this case. The fast-to-Alfv\'en conversion only occurs when the wave propagation is not contained in the same plane of the density stratification and magnetic field and, thus, three dimensional (3D) analysis are necessary. The study of this process was started by \citet{Crouch+Cally2005}, who studied the 3D propagation of oscillations in a polytrope permeated by an uniform magnetic field of arbitrary inclination, and found downward Alfv\'en waves. \citet{Cally+Goossens2008} obtained that the conversion to the Alfv\'en mode is most efficient for field inclinations from vertical between 30 and 40 degrees, and azimuth angles (the angle between the magnetic field and wave propagation planes) between 60 and 80 degrees. \citet{Cally+Hansen2011} found that the interaction between fast and Alfv\'en waves is spread across many scale height, unlike the fast-to-slow, which is limited around the layer where the sound speed $c_S$ and the Alfv\'en speed $v_A$ are similar. As the frequency increases the mode conversion is progressively more localized, although at the frequencis relevant to local helioseismology (around 3-5 mHz) the fast-to-Alfv\'en conversion region spans the whole chromosphere.       

Recently, Khomenko and Cally have studied the conversion to the Alfv\'en mode by means of 2.5D numerical simulations in homogeneous field configurations \citep{Khomenko+Cally2011}, as well as realistic sunspot-like structures \citep{Khomenko+Cally2012}. In these works they obtained the dependence of the efficiency of the conversion with the inclination and azimuth. However, in the sunspot configuration the conversion was only evaluated at some limited angles corresponding to selected 2D planes of the model. The aim of this work is to extend the results from \citet{Khomenko+Cally2012} to 3D. It will allow us to populate the voids in their diagrams of fast-to-Alfv\'en conversion efficiency. More relevant, the development of full 3D simulations provides the results for the complete wave field, where waves can propagate freely in all spatial directions without being restricted to a two-dimensional plane. The conversion to the Alfv\'en mode in realistic 3D atmospheres has not been studied before with the detailed evaluation of the efficiency discussed in this paper.

\section{Numerical procedures}
\label{sect:procedures}

The three-dimensional (3D) non-linear magnetohydrodynamic equations are solved using the numerical code Mancha \citep{Khomenko+Collados2006, Khomenko+etal2008, Felipe+etal2010a}. The code solves the equations for perturbations, obtained after removing the equilibrium state from the equations. A Perfect Matched Layer (PML) is used to avoid wave reflection at the top boundary \citep{Berenger1996}, while periodic boundary conditions are imposed in the horizontal boundaries. The initial perturbation is set in the bottom boundary with small amplitude in order to ensure that the simulations are in the linear regime.

As a background atmosphere we use a magnetostatic (MHS) sunspot model, adopted from \citet{Khomenko+Collados2008}. This model is a thick flux tube with distributed currents, azimuthally symmetric and has no twist. We set the height reference $z=0$ Mm at the photospheric level, where the optical depth at 500 nm is unity in the quiet Sun atmosphere. At this height the magnetic field at the axis is 900 G. The spatial resolution is 150 km in the horizontal directions and 50 km in the vertical direction. The computational domain spans from $z=-5$ Mm, where the minus sign indicates that it is below the photosphere, to $z=2.4$ Mm. The upper 500 km (10 grid points) correspond to the PML boundary layer, so the effective top of the simulation is located at $z=1.9$ Mm. The horizontal extent of the domain is $x \in [-39,39]$ Mm and $y \in [-30,30]$ Mm, with the axis of the sunspot located at $x=0$, $y=0$ Mm. At 39 Mm from the sunspot axis the thermodynamic variables of the model are taken from Model S \citep{Christensen-Dalsgaard+etal1996} in the deep sub-photosphere layers and VAL-C model \citep{Vernazza+etal1981} in the photosphere and chromosphere, stabilized following the method by \citet{Parchevsky+Kosovichev2007} to avoid the convective instability. The axis of the sunspot is given by the \citet{Avrett1981} model. The atmosphere between the quiet Sun boundary and the umbral model at the axis merges smoothly.

Waves are driven in a few grid points at the bottom boundary at $z=-5$ Mm. The perturbations in pressure, density, and velocity are calculated analytically as an acoustic-gravity wave of a given frequency and wavenumber, neglecting the magnetic field and temperature gradient \citep{Mihalas+Mihalas1984}. The detailed form of the perturbations can be found in \citet{Khomenko+Cally2012}. In this simulation the frequency was set to $\nu=\omega /2\pi=5$ mHz, slightly below the maximum cut off frequency reached at the temperature minimum, and the horizontal wave number to $k_x=1.37$ Mm$^{-1}$.

In order to indentify the slow, fast, and Alfv\'en wave modes in the magnetically dominated region of the computational domain, the velocity and magnetic field perturbations have been projected according to their orientation with respect to the equilibrium magnetic field onto these three characteristic directions:  

\begin{equation}
\label{eq:elong}
\hat{e}_{long}=({\rm cos}\phi {\rm sin}\theta, {\rm sin}\phi {\rm sin}\theta, {\rm cos}\theta),
\end{equation}
\begin{eqnarray}
\label{eq:eperp}
\lefteqn{\hat{e}_{perp}=(-{\rm cos}\phi {\rm sin}^2\theta {\rm sin}\phi,}\nonumber\\
&&1-{\rm sin}^2\theta {\rm sin}^2\phi,-{\rm cos}\theta {\rm sin}\theta {\rm sin}\phi)\,,
\end{eqnarray}
\begin{equation}
\label{eq:etrans}
\hat{e}_{trans}=(-{\rm cos}\theta,0,{\rm cos}\phi{\rm  sin}\theta).
\end{equation}

\noindent where $\theta$ is the magnetic field inclination from the vertical and $\phi$ is the field azimuth, measured from the $x-z$ plane. The projection $\hat{e}_{long}$ is along the magnetic field and it selects the slow magneto-acoustic wave in the low-$\beta$ region; the projection $\hat{e}_{perp}$ was chosen after \citet{Cally+Goossens2008} and it gives the asymptotic polarization direction of the Alfv\'en mode; and the last projection $\hat{e}_{trans}$ is set normal to the other two, corresponding to the fast wave in the low-$\beta$ regime. These projections have already been successfully used to separate the three wave modes in idealized magnetic field configurations \citep{Khomenko+Cally2011} as well as more complex magnetic topologies \citep{Felipe+etal2010a, Khomenko+Cally2012}. These projections assume that the wavevector ${\bf k}$ is contained in the $xz$ plane. In these simulations ${\bf k}$ will not be limited to that plane, since the raypath can bend due to background variations in the $y$ direction. This would have some effect in the accuracy of the projections, but it can probably be neglected.

As a measure of the efficiency of the conversion to each mode, the time-averaged wave energy fluxes \citep{Bray+Loughhead1974} were calculated in the magnetically dominated region. The acoustic energy flux is obtained from the expression:

\begin{equation}
{\bf F_{ac}}=\langle p_1{\bf v}\rangle,
\label{eq:Fac}
\end{equation}

\noindent while the magnetic energy flux is given by: 

\begin{equation}
{\bf F_{mag}}=\langle {\bf B_1}\times({\bf v}\times {\bf B_0})\rangle/\mu_0.
\label{eq:Fmag}
\end{equation}

\noindent where $p_1$, ${\bf v}$, and ${\bf B_1}$ are the Eulerian perturbations in pressure, velocity, and magnetic field, respectively, ${\bf B_0}$ is the background magnetic field, and $\mu_0$ is the magnetic permeability. In the region where $v_A>c_S$ the acoustic energy flux contains the energy of the slow mode, while the magnetic flux includes the fast and Alfv\'en modes. The time-averaged energy of the three wave modes was also calculated from the relations:

\begin{equation}
E_{long}=\rho_0c_S\langle v_{long}^2\rangle
\label{eq:Elong}
\end{equation}

\begin{equation}
E_{perp}=\rho_0v_A\langle v_{perp}^2\rangle
\label{eq:Eperp}
\end{equation}

\begin{equation}
E_{trans}=\rho_0v_A\langle v_{trans}^2\rangle
\label{eq:Etrans}
\end{equation}

\noindent where $v_{long}$, $v_{perp}$, and $v_{trans}$ are the velocity projections into the characteristic directions from Equations \ref{eq:elong}-\ref{eq:etrans}, and $\rho_0$ is the density in the equilibrium state. These expressions provide an approximation of the wave energy assuming equipartition between kinetic and other energies. In the case of pure acoustic or Alfv\'en waves, there is strict equipartition between kinetic and compressional or magnetic energy, respectively, and Equations (\ref{eq:Elong}-\ref{eq:Etrans}) correspond to the real energies.

\section{Velocity projections}
\label{sect:velocity}

When the fast acoustic wave which was driven at the bottom boundary reaches the $v_A=c_S$ layer several mode transformations take place. Above that layer, in the magnetically dominated atmosphere, the incident wave splits into a slow acoustic mode, a fast magnetic mode, and an Alfv\'en mode.

\begin{figure*}[!ht] 
 \centering
 \includegraphics[width=18cm]{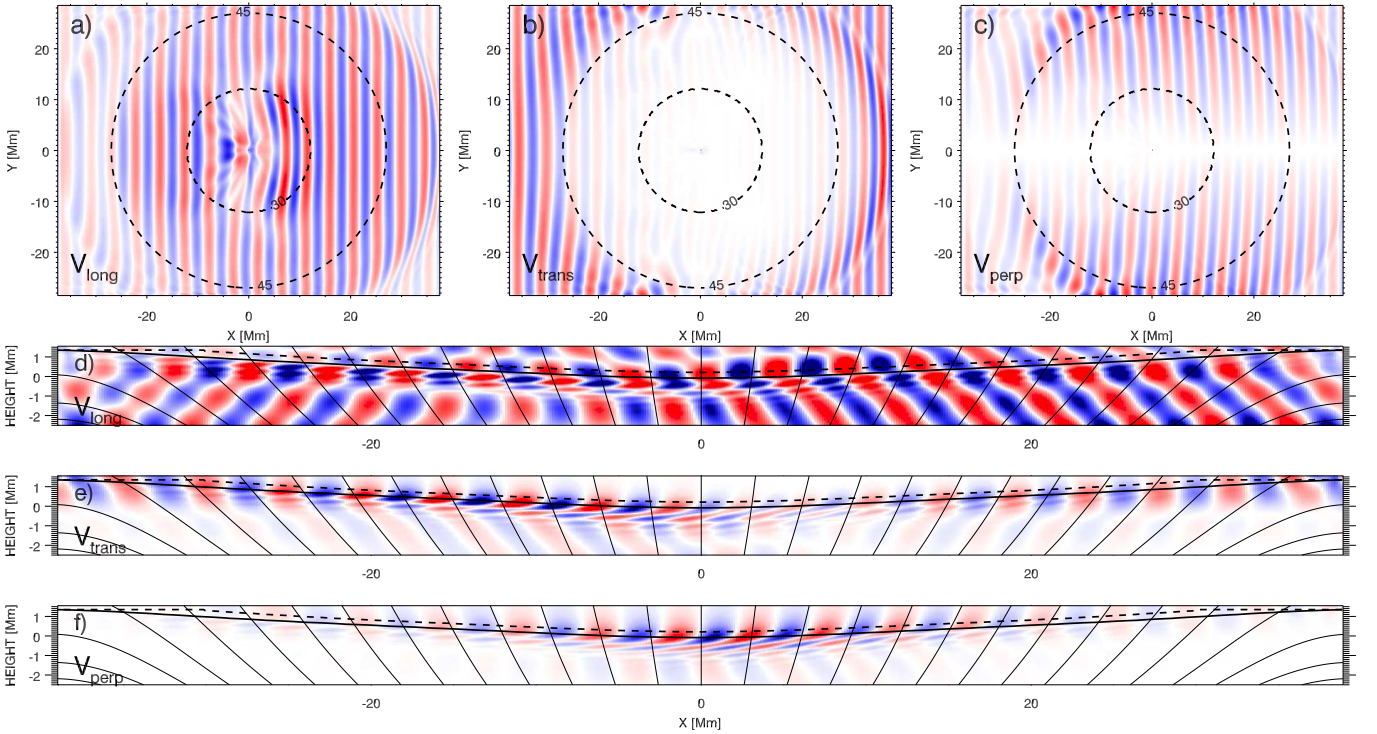}
  \caption{Snapshots of the three orthogonal components of the velocity at $t=19$ min. The blue-red colors mean positive-negative velocity directions; the range of the color coding is the same in all panels. The projection $v_{long}$ is scaled with a factor of $\sqrt{\rho_0c_S}$ and the projections $v_{trans}$ and $v_{perp}$ with a factor $\sqrt{\rho_0v_A}$. Top panels correspond to horizontal cuts at $z=1.65$ Mm, from left to right: $v_{long}$, $v_{trans}$, and $v_{perp}$. Dashed lines represents contours of equal inclination of the background magnetic field, measured at the height where $c_S=v_A$. The three bottom panels show vertical cuts at $y=7.5$ Mm, from top to bottom: $v_{long}$, $v_{trans}$, and $v_{perp}$. Horizontal solid line is the height where $c_S=v_A$; horizontal dashed line is the fast mode reflection level. Magnetic field lines are inclined black lines.}
  \label{fig:velocities}
\end{figure*}

Figure \ref{fig:velocities} shows snapshots of the projected velocities scaled with a factor $\sqrt{\rho_0v_{ph}}$, where $v_{ph}=c_S$ for the $v_{long}$ component and $v_{ph}=v_A$ for the other two components, after 19 min of simulation. The stationary regime is achieved after about 10 minutes of simulations. The inclination of the magnetic field $\theta$ varies from $0^o$ at the center of the sunspot to being almost horizontal at the boundaries of the computatial domain. The azimuth $\phi=0^o$ corresponds to all the positions at $y=0$ Mm with positive $X$ value, and it increases up to $\phi=180^o$ at $y=0$ Mm and negative $X$ values, including all the angles between these two extremes.

Each of the waves modes presents a different distribution across the 3D atmosphere of the sunspot. In the magnetically dominated atmosphere the slow wave appears in the $v_{long}$ projection. The $xy$ cut (top left panel) shows that the conversion to the slow mode is significative at almost all the positions of the sunspot. However, there are some locations where this transformation is specially favoured. The strongest slow wave signal appears for $X$ between $3$ and $12$ Mm and $Y$ between $-10$ and $10$ Mm. This region corresponds to moderate inclinations around $\theta =30^o$. Although the frequency of the wave is below the cut-off frequency, the slow mode can reach the upper atmosphere because of the reduced cut-off value due to the inclination of the magnetic field. However, the vertical magnetic field at the axis of the sunspot avoids the propagation of slow waves around the center of the sunspot, and they produce evanescent modes. In general, the amplitude of the slow mode in the x-positive region of the sunspot is higher than the x-negative region. The driving perturbation generates an acoustic-gravity wave which propagates from left to right. Thus, the right half of the sunspot has a better alignment between the direction of propagation and the field lines, producing a more efficient conversion from fast acoustic waves (in the region where $v_A<c_S$) to slow acoustic waves (in the region where $v_A>c_S$).        

\begin{figure}[!ht] 
 \centering
 \includegraphics[width=8.5cm]{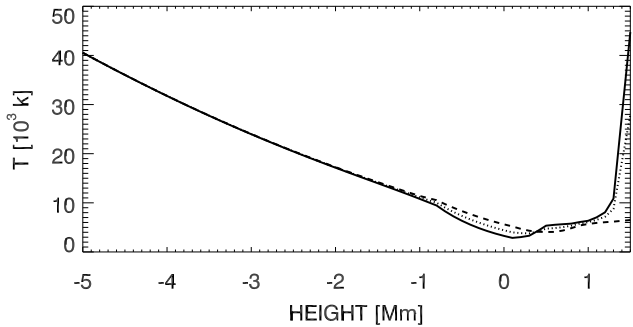}
  \caption{Variation of the temperature with height at the axis of the sunspot (solid line), at 7.5 Mm from the axis (dotted line), and at the quiet Sun atmosphere (dashed line). }
  \label{fig:temperature}
\end{figure}

The vertical cut of the slow mode (Figure \ref{fig:velocities}d) shows an interference pattern above the solid line, near the top boundary. It appears mainly in the left part of the sunspot, between $x=-20$ Mm and $x=5$ Mm. The interference is produced between the upward propagating slow mode and a downward propagating wave produced by the partial reflection of the slow mode due to the steep temperature increase at the chromosphere. Similar result was found in the 2.5D simulations from \citet{Khomenko+Cally2012}, where they confirmed that this reflection is a physical effect rather than an artifact from the top boundary. This reflection is stronger near the center of the sunspot, where the gradient of the temperature is steeper (Figure \ref{fig:temperature}), as can be seen in panel $(a)$ from Figure \ref{fig:velocities}. When the reflected wave reaches the $c_S=v_A$ layer it undergoes a secondary transformation, generating a new fast acoustic mode and a slow magnetic mode visible below the solid line in Figures \ref{fig:velocities}d and \ref{fig:velocities}e, respectively.

The fast magnetic mode in the low-$\beta$ region is reflected back down due to the gradients of the Alfv\'en speed (Figure \ref{fig:velocities}e). If we neglect the contribution of the sound speed to the fast wave speed, the reflection height is given by the layer where the wave frequency $\omega$ and the horizontal wave number $k_x$ are related by $\omega=v_Ak_x$, and it is represented in the figure by a dashed line. Around the center of the sunspot the reflection is completed and at the height where panel $(b)$ is obtained there is no fast mode in that region. Farther from the axis of the sunspot the transformation layer is located at a higher height, and the insufficient height of the top boundary of the computational domain avoids the complete reflection of the fast mode. 

The Alfv\'en wave can be seen in panels $(c)$ and $(f)$. Along $y=0$ Mm, which corresponds to $\phi=0^o$, there is no wave power in the Alfv\'en mode. At this position the magnetic field is contained in the $xz$ plane, since $B_y=0$ G, making this plane equivalent to a 2D case. Under these conditions the Alfv\'en mode is decoupled from the fast and slow magneto-acoustic modes, and it is not possible for the incident wave to undergo conversion to the Alfv\'en mode. As regions farther from the $y=0$ Mm plane are considered the conversion to the Alfv\'en mode becomes more efficient. It shows a highest amplitude at around $\theta=45^o$.

\begin{figure*}[h]
\centering
\subfloat{
\includegraphics[width=0.33\textwidth]{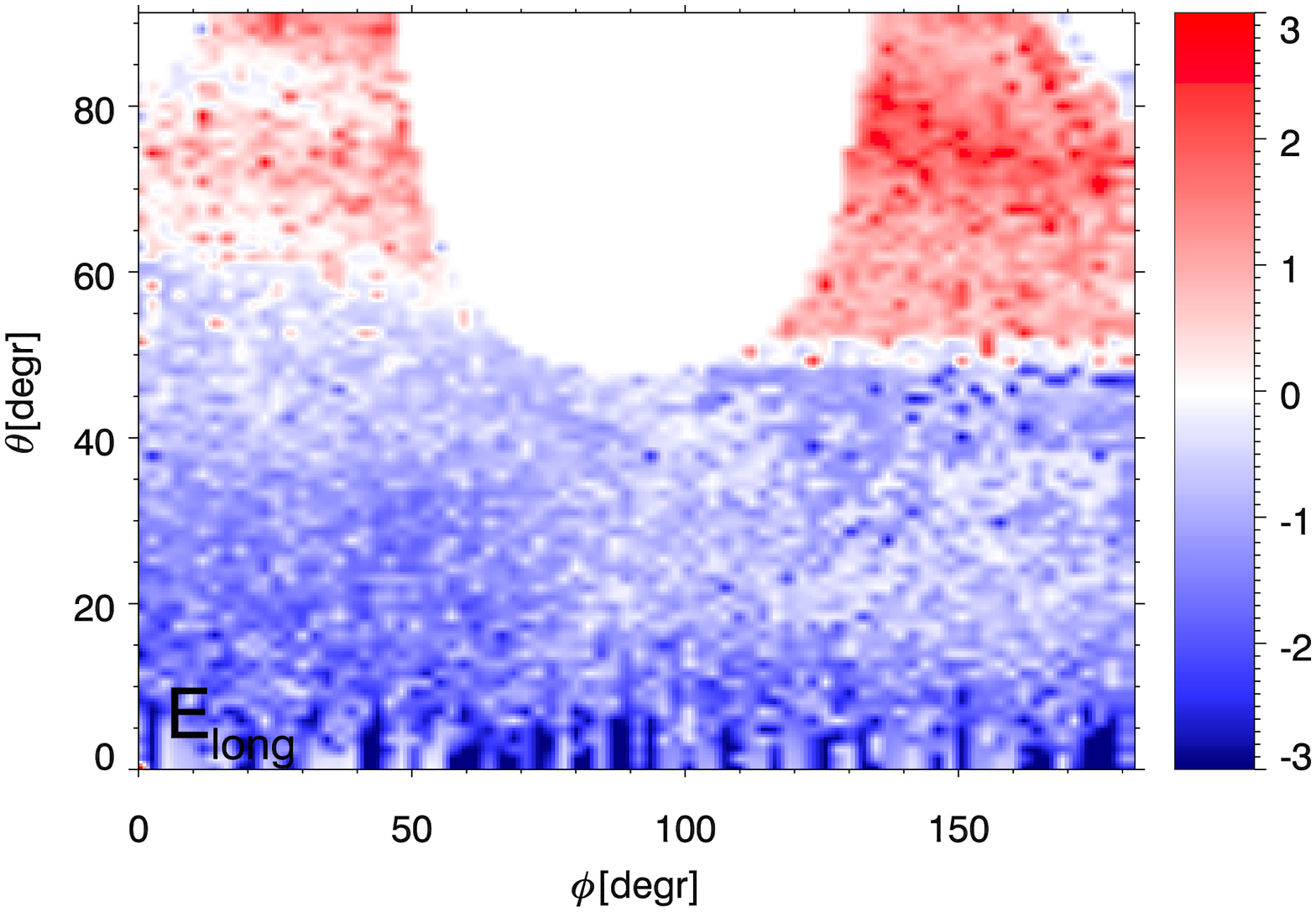}}
\subfloat{
\includegraphics[width=0.33\textwidth]{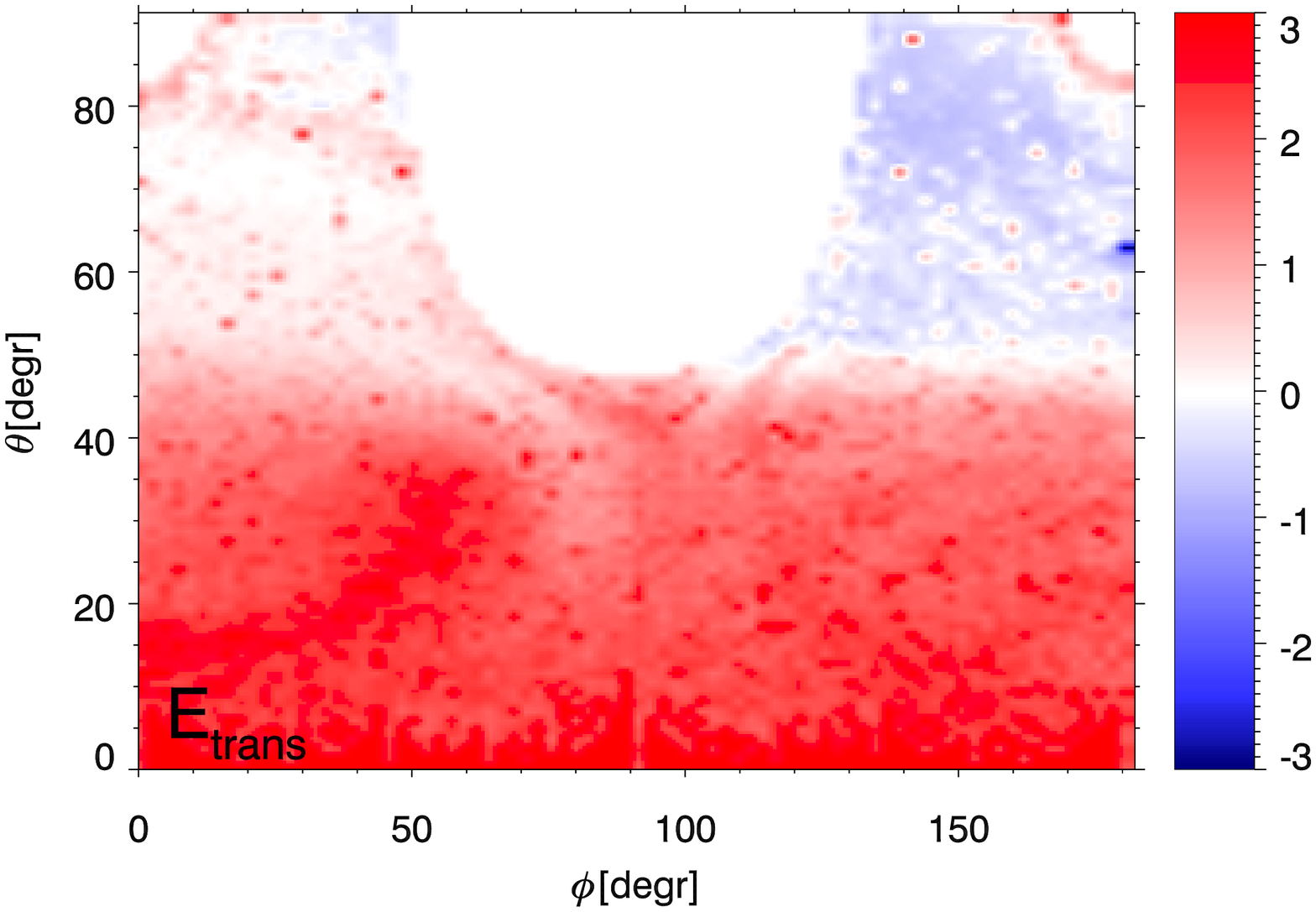}}
\subfloat{
\includegraphics[width=0.33\textwidth]{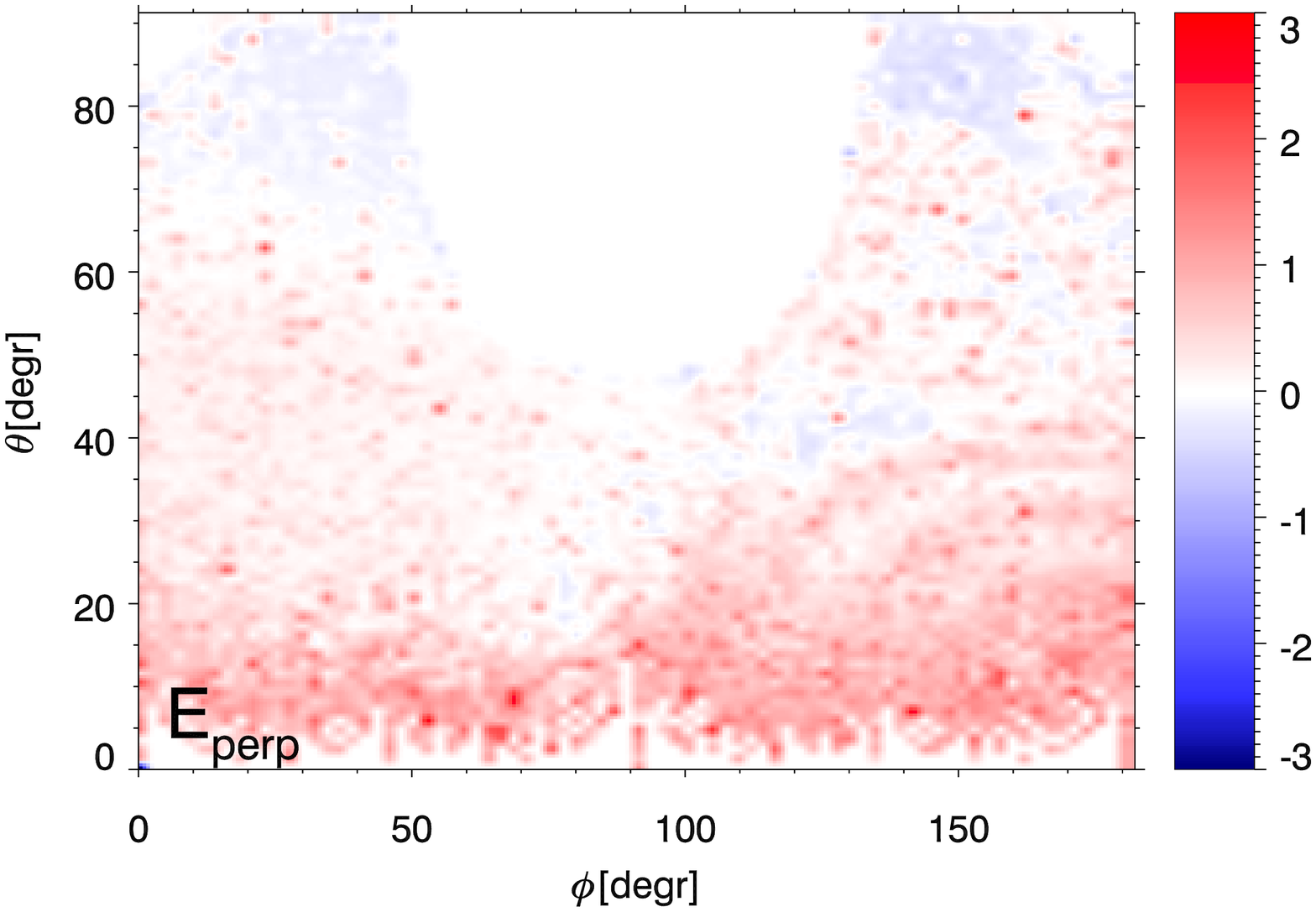}}
\caption[width=\textwidth]{Log$_{10}$ of the amplitude ratio $R={\bf B_1}/\sqrt{\mu_0\rho_0}/{\bf v_1}$ for the projected directions at the upper atmosphere (averages at heights from 0.4 Mm above the $c_S = v_A$ layer up to z = 1.9 Mm). Left panel: slow acoustic mode ($\hat{e}_{long}$); middle panel: fast magnetic mode ($\hat{e}_{trans}$); right panel: Alfv\'en mode ($\hat{e}_{perp}$)}
\label{fig:polarization}
\end{figure*}

\section{Alfv\'en mode polarization relations}
\label{sect:polarization}

Before evaluating the contribution of the different wave modes to the energy in the upper part of the atmosphere, we have checked the validity of the projections described in the previous section, in order to ensure that they provide an accurate decoupling of the slow, fast, and Alfv\'en waves. Following \citet{Khomenko+Cally2012}, we have calculated the ratio $R={\bf B_1}/\sqrt{\mu_0\rho_0}/{\bf v_1}$. Since the kinetic and magnetic energy for an Alfv\'en wave is in equipartition, one would expect the ratio $R$ to be equal to one for that waves.

Figure \ref{fig:polarization} shows the ratio $R$ for the three projected components of the velocity and magnetic field. In each case, ${\bf B_1}$ and ${\bf v_1}$ pairs were obtained from the decomposition in the directions defined by Equations \ref{eq:elong} and \ref{eq:etrans}. The ratio is evaluated at heights from 400 km above the transformation layer to the top boundary, and it is averaged in time for the stationary stage of the simulations. It reveals the different nature of the three projections. For the $\hat{e}_{long}$ direction at $\theta <50^o$, the ratio $R$ present small values around $10^{-2}$, indicating that this component is dominated by the velocity variations rather than magnetic, and confirming that this projection contains the slow acoustic waves. However, at higher inclinations the magnetic variations associated with velocity variations are bigger. An opposite behavior is found for the $\hat{e}_{trans}$ component. In this case, in the regions where $\theta <50^o$ the ratio $R$ is big, around $10^2-10^3$, and it becomes smaller at higher inclinations. These plots indicate that the projections $\hat{e}_{long}$ and $\hat{e}_{trans}$ provide a good estimation of the slow and fast waves, respectively, in the upper atmosphere when the inclination of the magnetic field is below $50^o$. At higher inclinations both wave modes are mixed up, since the projection directions are asymptotic, valid strictly only where the Alfv\'en speed is much higher than the sound speed. At these locations this criterion is not fulfilled.

The right panel of Figure \ref{fig:polarization} illustrates the ratio $R$ for the $\hat{e}_{perp}$ component.  It shows a ratio around $10^{0}$ for all the atmosphere, indicating that for this projections the velocity and magnetic perturbations are in equipartition, which confirm the Alfv\'enic nature of the oscillations in this characteristic direction.

\begin{figure*}[h]
\centering
\subfloat{
\includegraphics[width=0.66\textwidth]{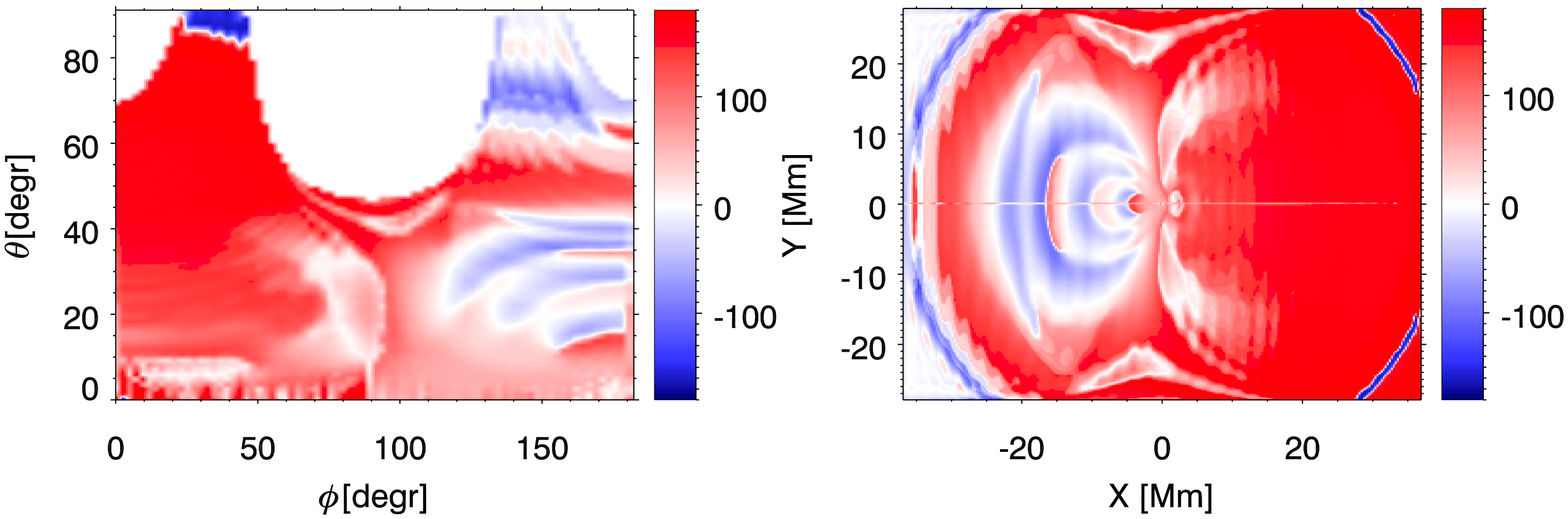}}
\qquad
\subfloat{
\includegraphics[width=0.66\textwidth]{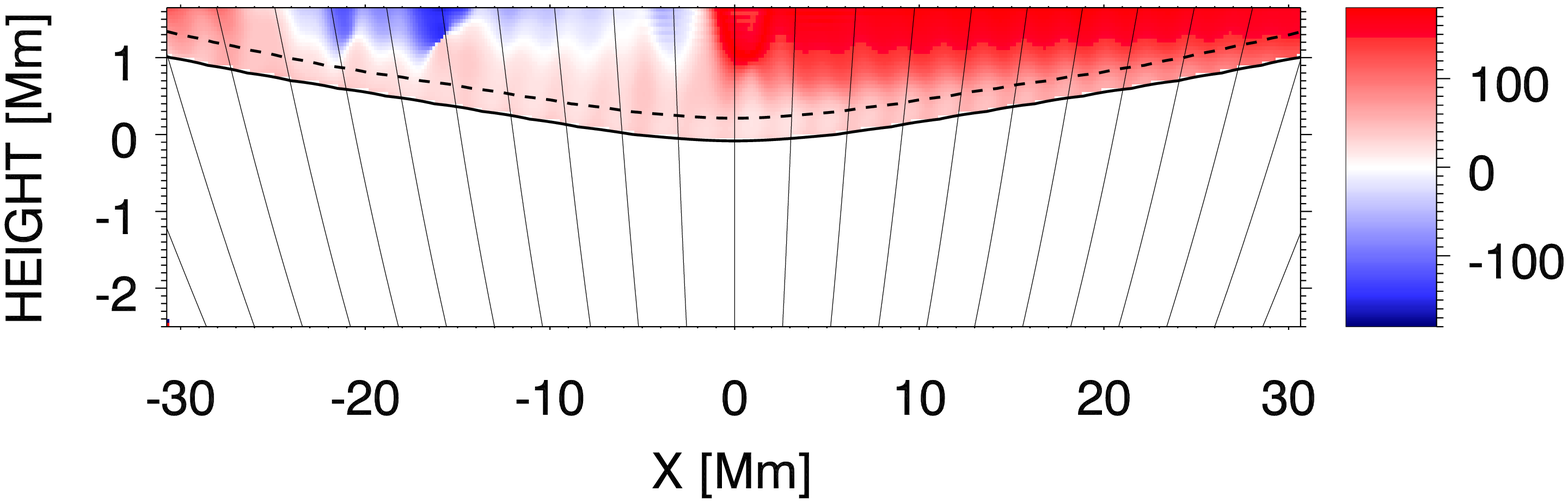}}
\caption[width=\textwidth]{Phase shift between the variations of $v_{perp}$ and $B_{perp}$  in the upper atmosphere (averages at heights from 0.4 Mm above the $c_S = v_A$ layer up to z = 1.9 Mm) as a function of inclination and azimuth (top left panel) and X and Y (top right panel). Bottom panel shows the same quantity, but as a function of height and horizontal distance at $y=7.5$ Mm.}
\label{fig:phase}
\end{figure*}

Figure \ref{fig:phase} shows the phase difference between the perturbations in the velocity and magnetic field in the direction $\hat{e}_{perp}$, which correspond to the Alfv\'en wave. For a pure Alfv\'en mode a phase shift of 180$^o$ indicates upward propagation, while negative phase shifts correspond to downward propagation. Most of the right part of the sunspot, that is, for azimuths between $0^o$ and $90^o$, presents a phase shift around $180^o$. At these locations the Alfv\'en waves come from the conversion of the upward propagating fast acoustic mode introduced in the high-$\beta$ region, and they keep their upward propagation to higher layers. However, at some positions in the left side of the sunspot (with $\phi$ between $90^o$ and $180^o$) the negative sign of the phase shift indicates downward propagating waves. The direction of the propagation of the Alfv\'en waves in this simulation shows a perfect agreement with the one obtained by \citet{Khomenko+Cally2012}. Since the efficiency of the conversion to the Alfv\'en mode is enhanced with the alignment of the direction of propagation and magnetic field, in the right part of the sunspot the upward propagating fast waves couple to upward Alfv\'en waves. On the other hand, in the left part of the sunspot the most efficient conversion to Alfv\'en waves occurs for the refracted downward propagating fast waves \citep{Cally+Hansen2011}. See Figure 1 from \citet{Khomenko+Cally2012} for a schematic representation of these mode transformations.

\section{Energy fluxes}
\label{sect:flux}

\begin{figure*}[!ht] 
 \centering
 \includegraphics[width=18cm]{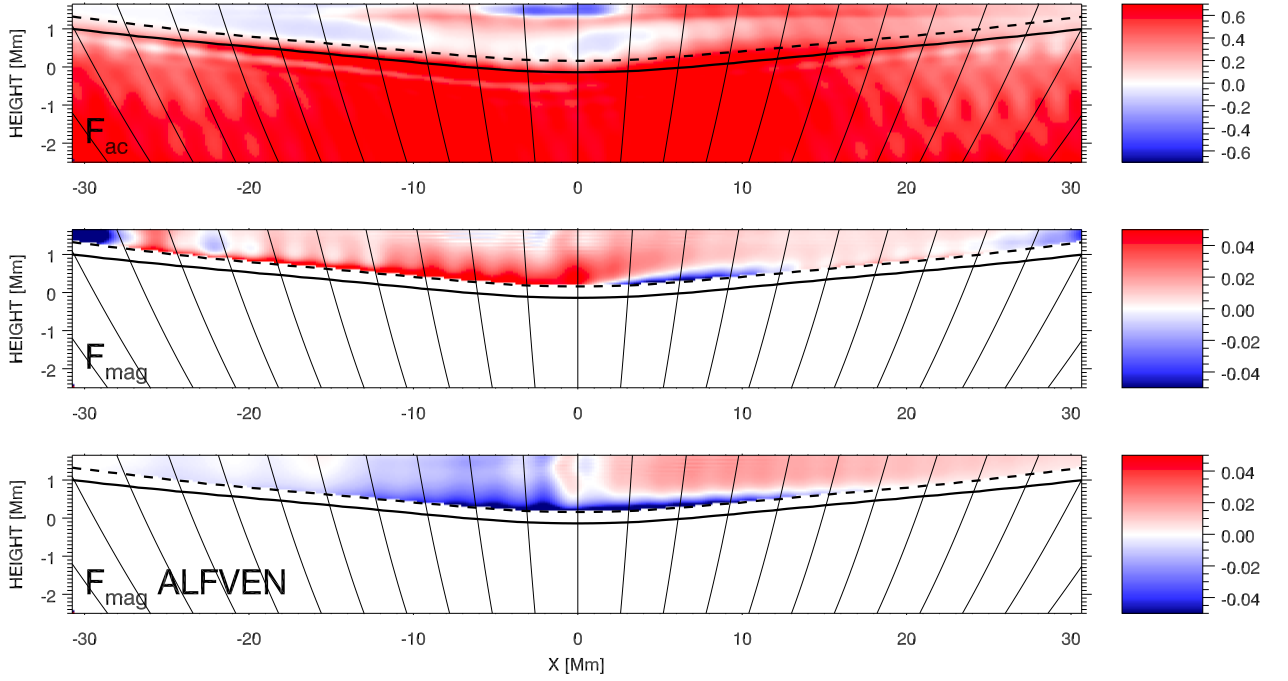}
  \caption{Vertical component of the energy fluxes at $y=6$ Mm. Top panel: acoustic flux; middle panel: magnetic flux; bottom panel: magnetic flux due to Alfv\'en waves. The units of the color coding are $10^6$ erg cm$^{-2}$s$^{-1}$. Positive fluxes mean energy propagating upward and negative fluxes downward. Horizontal solid line is the height where $c_S=v_A$; horizontal dashed line is the fast mode reflection level. Magnetic field lines are inclined black lines.}
  \label{fig:fluxXZ}
\end{figure*}

\begin{figure*}[h]
\centering
\subfloat{
\includegraphics[width=0.33\textwidth]{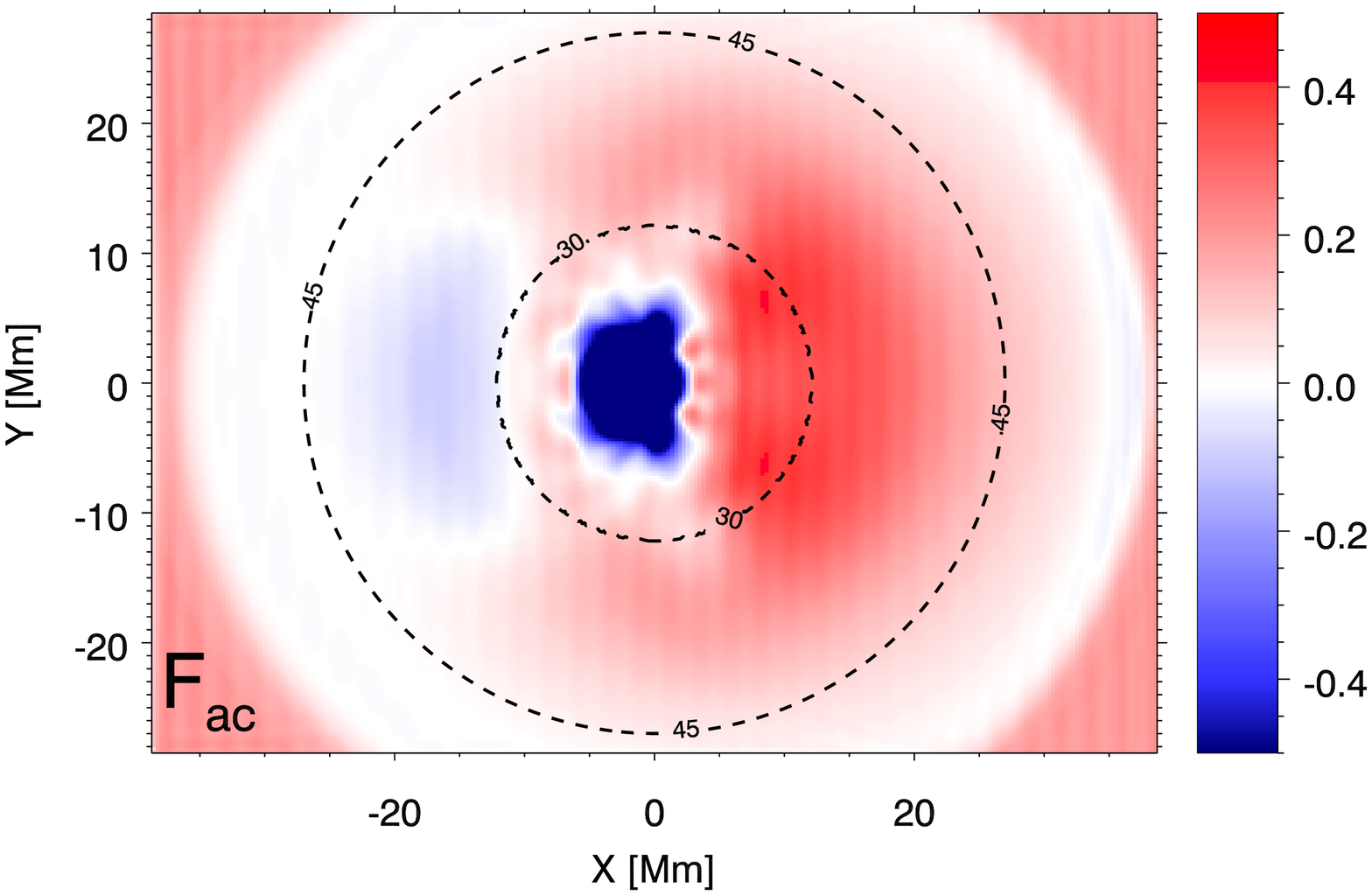}}
\subfloat{
\includegraphics[width=0.33\textwidth]{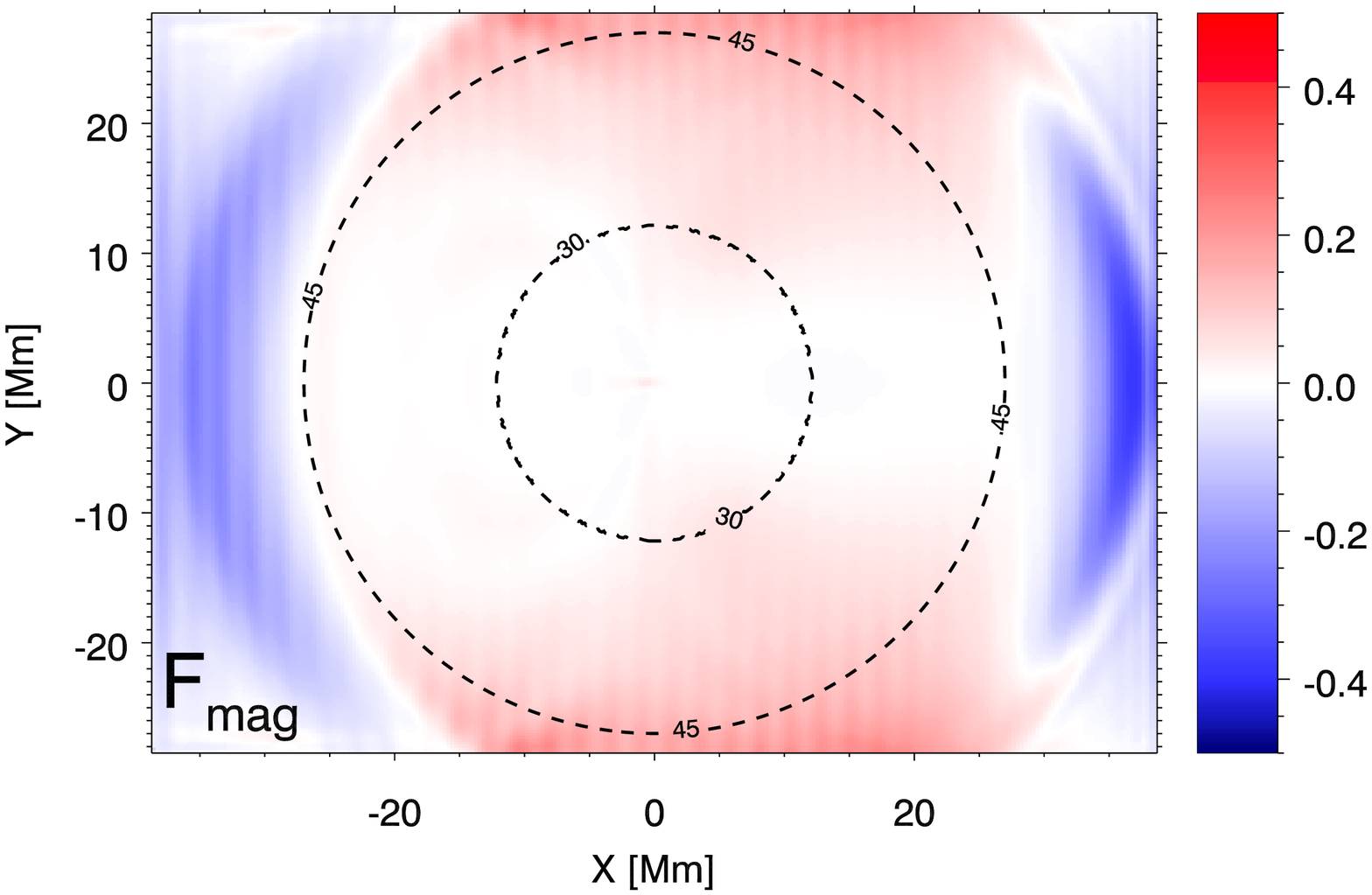}}
\subfloat{
\includegraphics[width=0.33\textwidth]{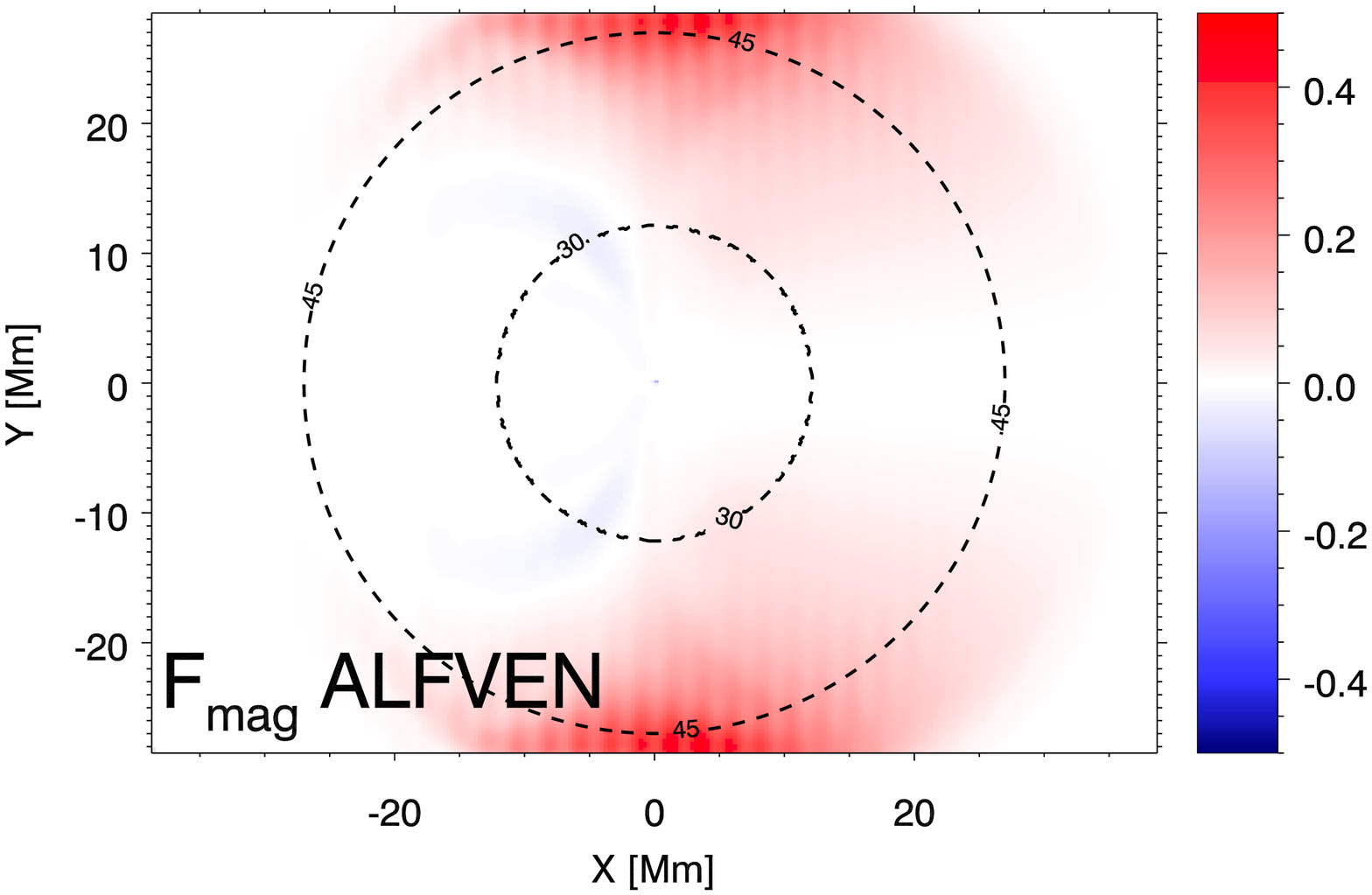}}
\caption[width=\textwidth]{Vertical component of the energy fluxes at $z=1.65$ Mm. Left panel: acoustic flux; middle panel: magnetic flux; right panel: magnetic flux due to Alfv\'en waves. The units of the color coding are $10^6$ erg cm$^{-2}$s$^{-1}$. Positive fluxes mean energy propagating upward and negative fluxes downward.}
\label{fig:fluxXY}
\end{figure*}

The acoustic and magnetic energy fluxes were calculated using Equations \ref{eq:Fac} and \ref{eq:Fmag}, respectively. Figures \ref{fig:fluxXZ} and \ref{fig:fluxXY} show the time-average results, including all the time steps after the stationary regime is achieved. The former one corresponds to a $xz$ cut at $y=6$ Mm. The magnetic flux is only plotted above the $c_S=v_A$ layer, where the distinction of the three different modes using the projections described in Equations \ref{eq:elong}-\ref{eq:etrans} is meaningful. The later figure shows a $xy$ plot at $z=1.65$ Mm.

Most of the domain presents a positive acoustic flux, which corresponds to the upward propagating slow mode in the low-$\beta$ region. The highest contribution of this mode is located at the right side of the sunspot, especially at moderate inclinations where the amplitude of the $v_{long}$ projections was higher in Figure \ref{fig:velocities}. However, at the left part of the sunspot a negative flux is obtained. This downward propagating slow mode flux is particularly large near the axis of the sunspot and represents the slow waves reflected by the temperature gradient, as discussed in the previous section. The 5 mHz frequency fast acoustic wave which propagates upward in the high-$\beta$ region reaches the conversion layer $c_S=v_A$ before the high value of the cut-off frequency avoids its propagation. Note that near the center of the sunspot the cut-off frequency presents its highest value of $5.7$ mHz at the temperature minimum, which is located 375 km above the $c_S=v_A$ layer. Just above this layer, the recently converted slow acoustic mode becomes an evanescent wave. This situation differs from what happen at locations far from the axis of the sunspot, where the field inclination reduces the effective cut-off and allows the slow mode waves to scape to the upper atmosphere. At a height located a few hundreds kilometers higher than the temperature minimum, the cut-off frequency of the atmosphere is again below 5 mHz due to the chromospheric increase of the temperature, and the slow mode near the axis can propagate again, allowing the downward propagation of the waves reflected by the rise of the temperature.

Above the transformation layer the magnetic energy flux includes the fast and Alfv\'en modes. The middle panel of Figure \ref{fig:fluxXZ} shows that the magnetic flux is positive at most heights, while in the middle panel of Figure \ref{fig:fluxXY} can be seen how it increases in the $Y$ direction toward the periphery of the sunspot. On the other hand, near the $X$ boundaries of the model a negative magnetic flux appears, larger than the highest positive magnetic flux. This negative flux must be produced by the reflected fast wave.

For a more detailed analysis of the contribution of the different wave modes, the magnetic flux of the Alfv\'en wave was deattached from the total magnetic flux by recalculating the magnetic flux but using the projections of the velocity and magnetic field in the $\hat{e}_{trans}$ direction in Equation \ref{eq:Fmag}. The result is plotted in the bottom and right panels of Figures \ref{fig:fluxXZ} and \ref{fig:fluxXY}, respectively. The vertical cut reveals that at $y=6$ Mm the vertical Alfv\'en flux is positive in the righ part and negative in the left part, that is, at the right part of the domain the Alfv\'en wave propagate upward and at the left part they propagate downward. This result agrees with the one obtained in the previous section from the phase shift between $v_{perp}$ and $B_{perp}$. The $xy$ cut shows that the Alfv\'en energy flux increases at locations with higher inclinations, far from the axis of the sunspot, only for those regions of the atmosphere whose azimuth $\phi$ is different from $0^o$. Thus, the highest Alfv\'en energy flux is obtained near the $Y$ boundaries. Note that it is comparable to the highest acoustic energy flux. The negative flux at the left side of the sunspot shown in the bottom panel of Figure \ref{fig:fluxXZ} is hardly visible in the right panel of Figure \ref{fig:fluxXY} because of the different scale used in the plots. 

A comparison between middle and right panels of Figure \ref{fig:fluxXY} reveals that the negative magnetic flux from the $X$ boundaries is missing in the Alfv\'en energy flux, so it corresponds to the reflected fast mode, as previously stated. At the positions where the Alfv\'en energy flux is larger the total magnetic flux shows a lower positive value, meaning that at these locations the fast wave must also contribute with a negative flux.

\section{Energy of the three wave modes}
\label{sect:energy}

The wave energy of the slow, fast, and Alfv\'en modes was calculated from Equations \ref{eq:Elong}-\ref{eq:Etrans}, using the corresponding projected velocities. To be consistent with the work by \citet{Khomenko+Cally2012} and provide a direct comparison with their 2.5D simulations, time averaged energies were obtained at heights from 400 km above the $c_S=v_A$ layer up to the upper boundary of the simulation box in the stationary stage of the simulations.

\begin{figure*}[h]
\centering
\subfloat{
\includegraphics[width=0.33\textwidth]{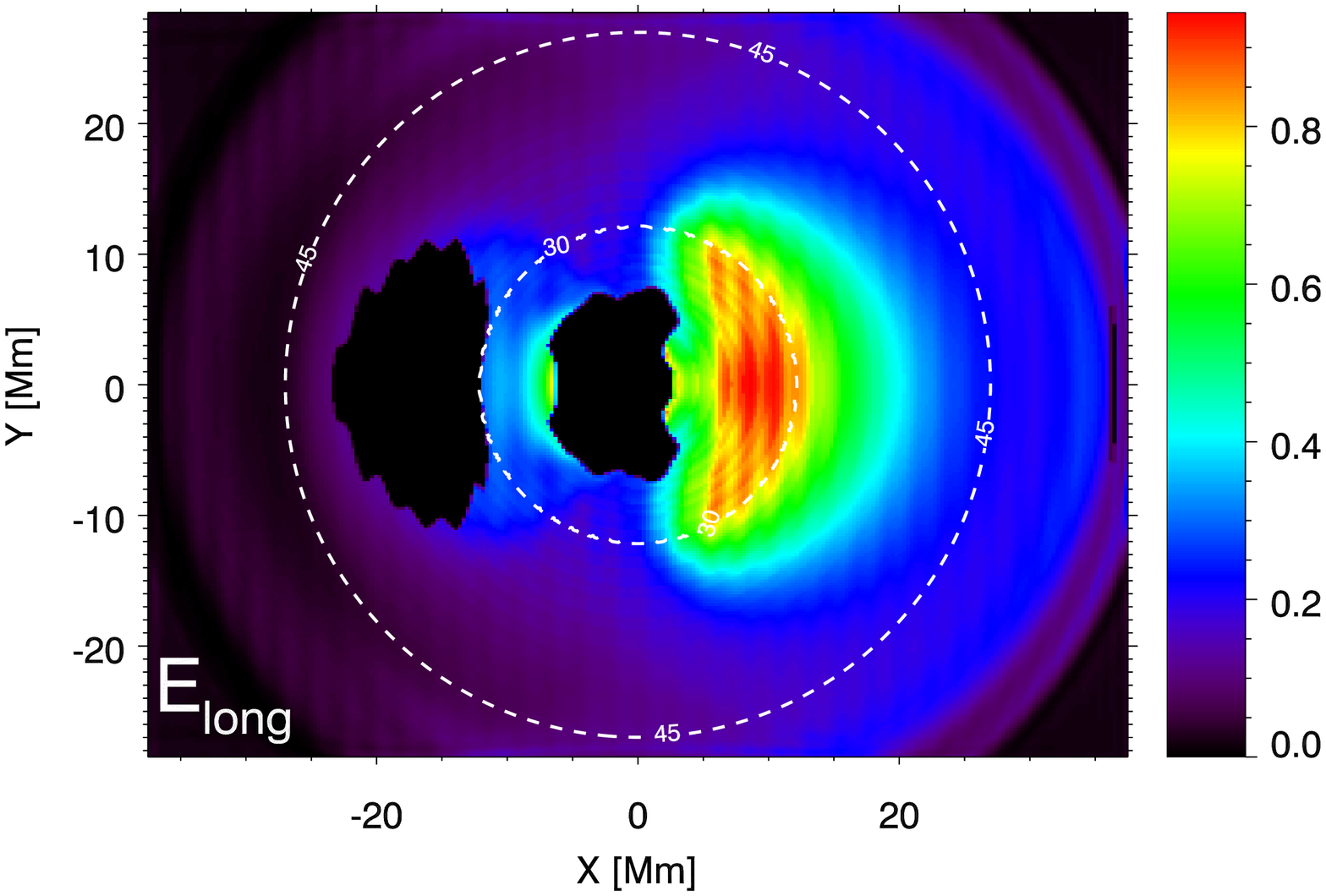}}
\subfloat{
\includegraphics[width=0.33\textwidth]{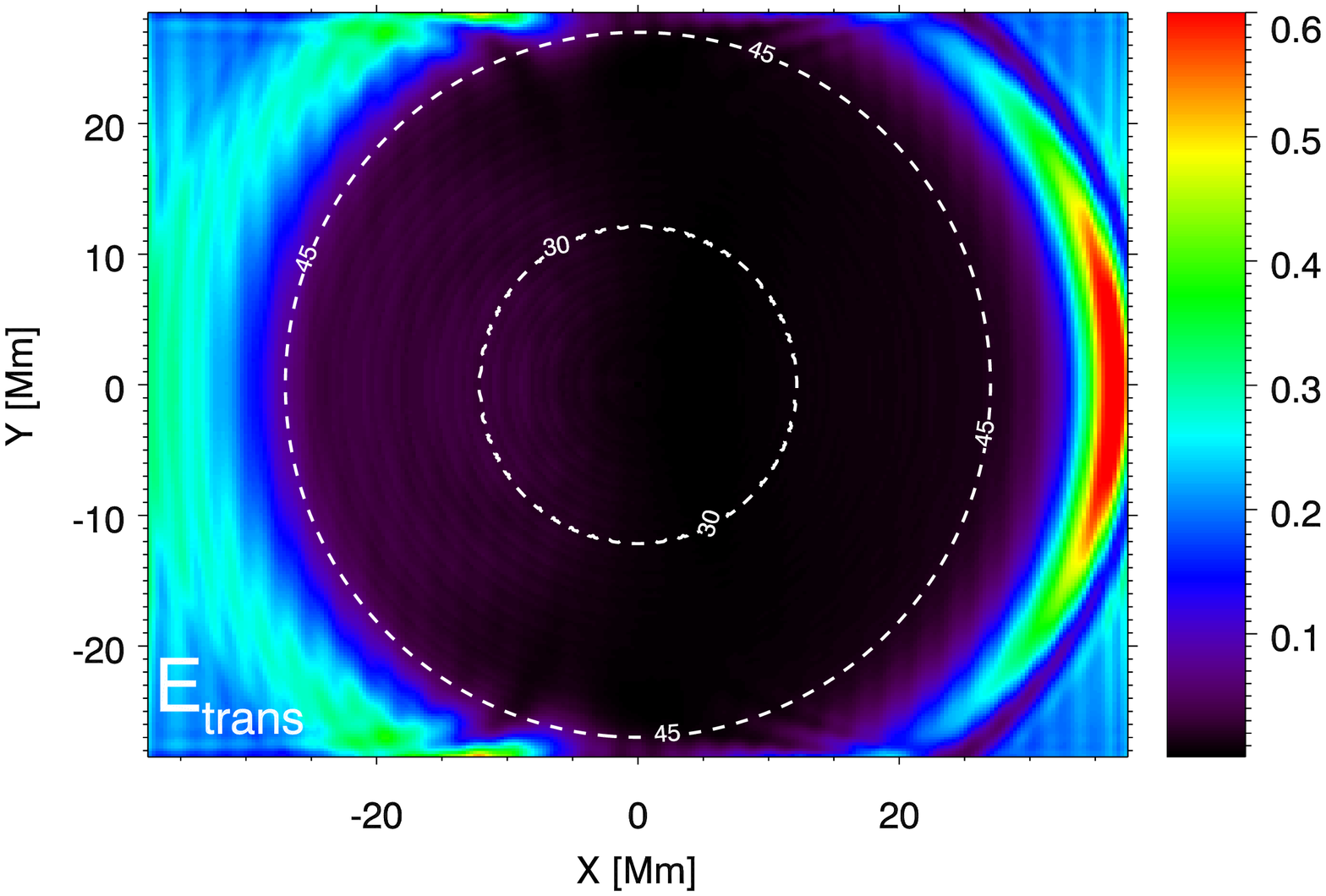}}
\subfloat{
\includegraphics[width=0.33\textwidth]{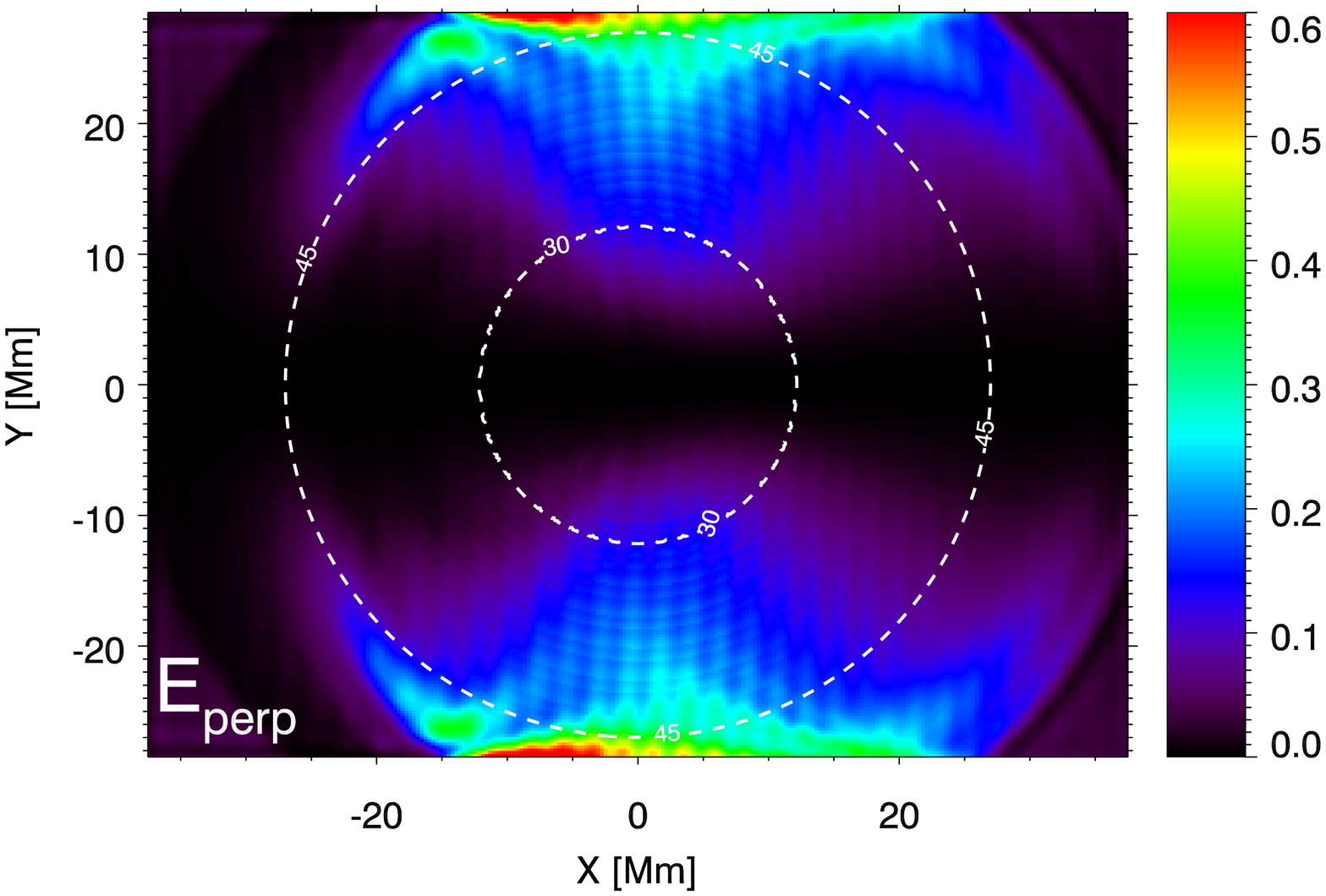}}
\caption[width=\textwidth]{Wave energies at the top of the atmosphere (averages
at heights from 0.4 Mm above the $c_S = v_A$ layer up to z = 1.9 Mm) for the three projected velocity components as a function of the horizontal locations. Left panel: acoustic flux; middle panel: magnetic flux; right panel: magnetic flux due to Alfv\'en waves. The units of the color coding are $10^6$ erg cm$^{-2}$s$^{-1}$. In the left panel the regions with negative acoustic vertical flux have been masked.}
\label{fig:energy_XY}
\end{figure*}

Figure \ref{fig:energy_XY} illustrates the results for all the horizontal positions from the computational domain. The left panel shows that there is a prominent maximum of the slow wave energy at the right part of the sunspot, not far from the axis. The highest slow mode energy apears at around $x=10$ Mm and $y=0$ Mm. The main interest of this measurement is to quantify the amount of energy which is converted at the $c_S=v_A$ from the incident fast acoustic mode to the outgoing upward propagating slow acoustic mode. For this reason, in the left panel the regions with negative acoustic flux in Figure \ref{fig:fluxXY} have been masked, since the slow wave energy in that locations does not comes directly from the conversion layer but from the reflection due to the temperature gradient.

\begin{figure*}[h]
\centering
\subfloat{
\includegraphics[width=0.33\textwidth]{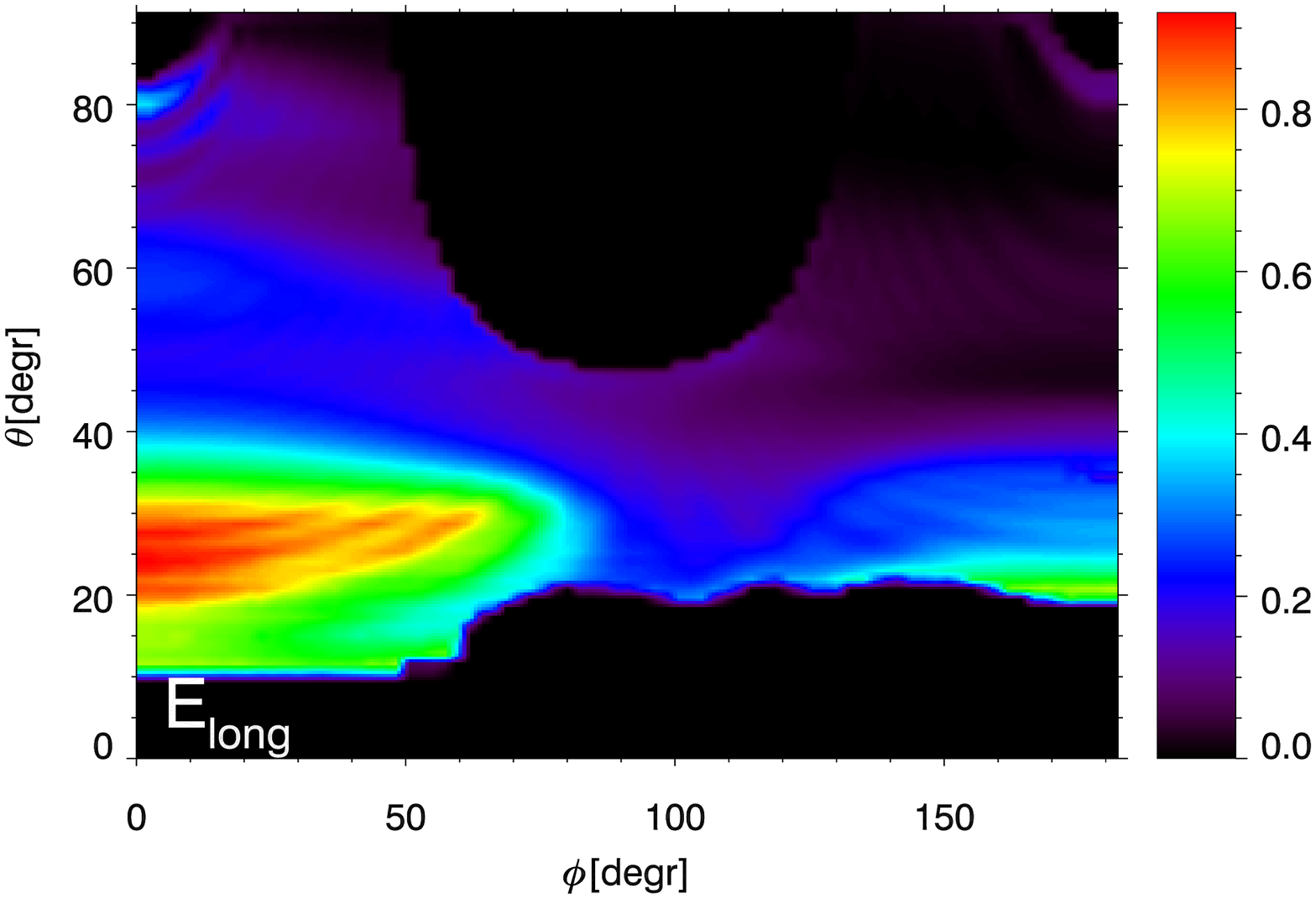}}
\subfloat{
\includegraphics[width=0.33\textwidth]{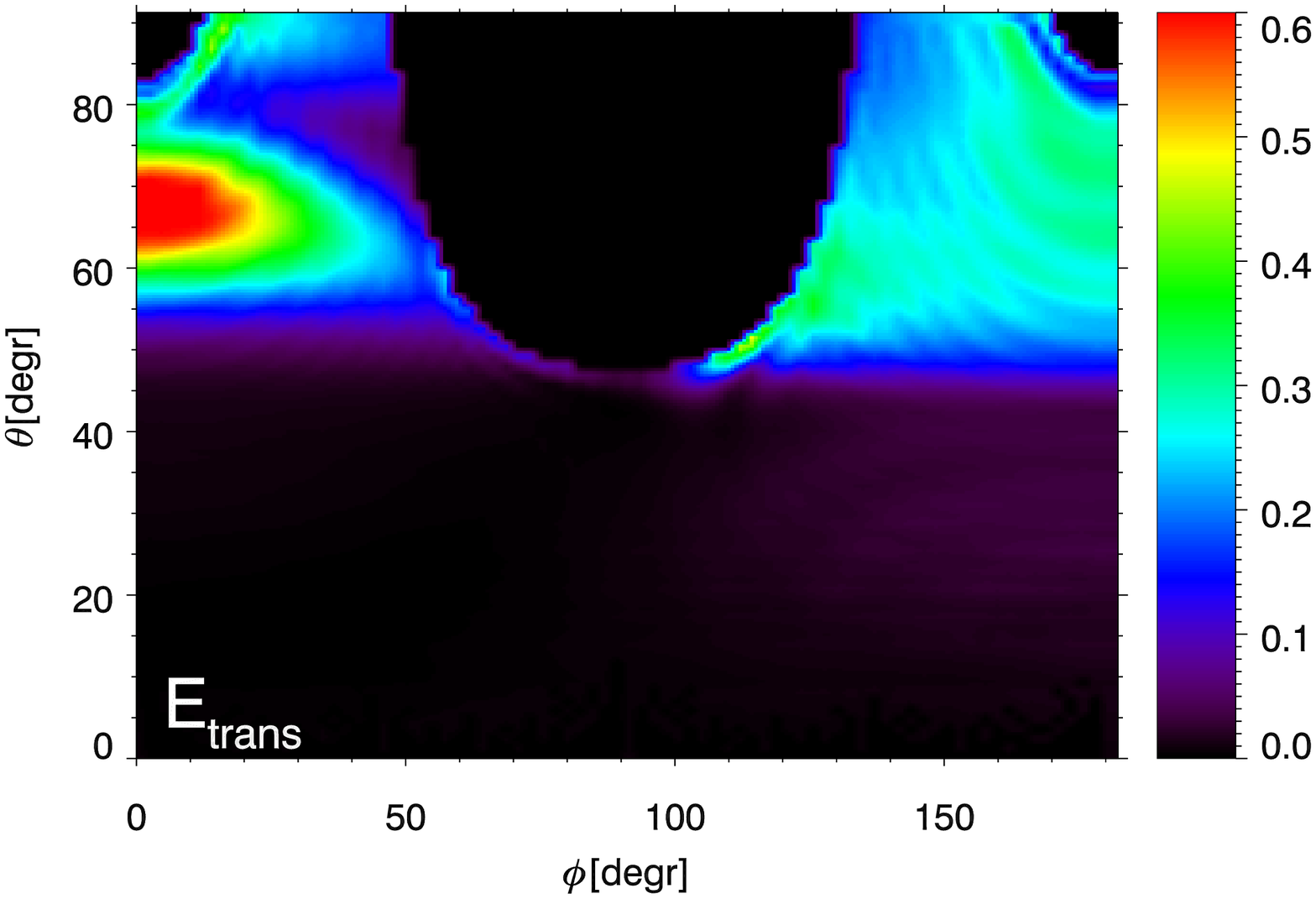}}
\subfloat{
\includegraphics[width=0.33\textwidth]{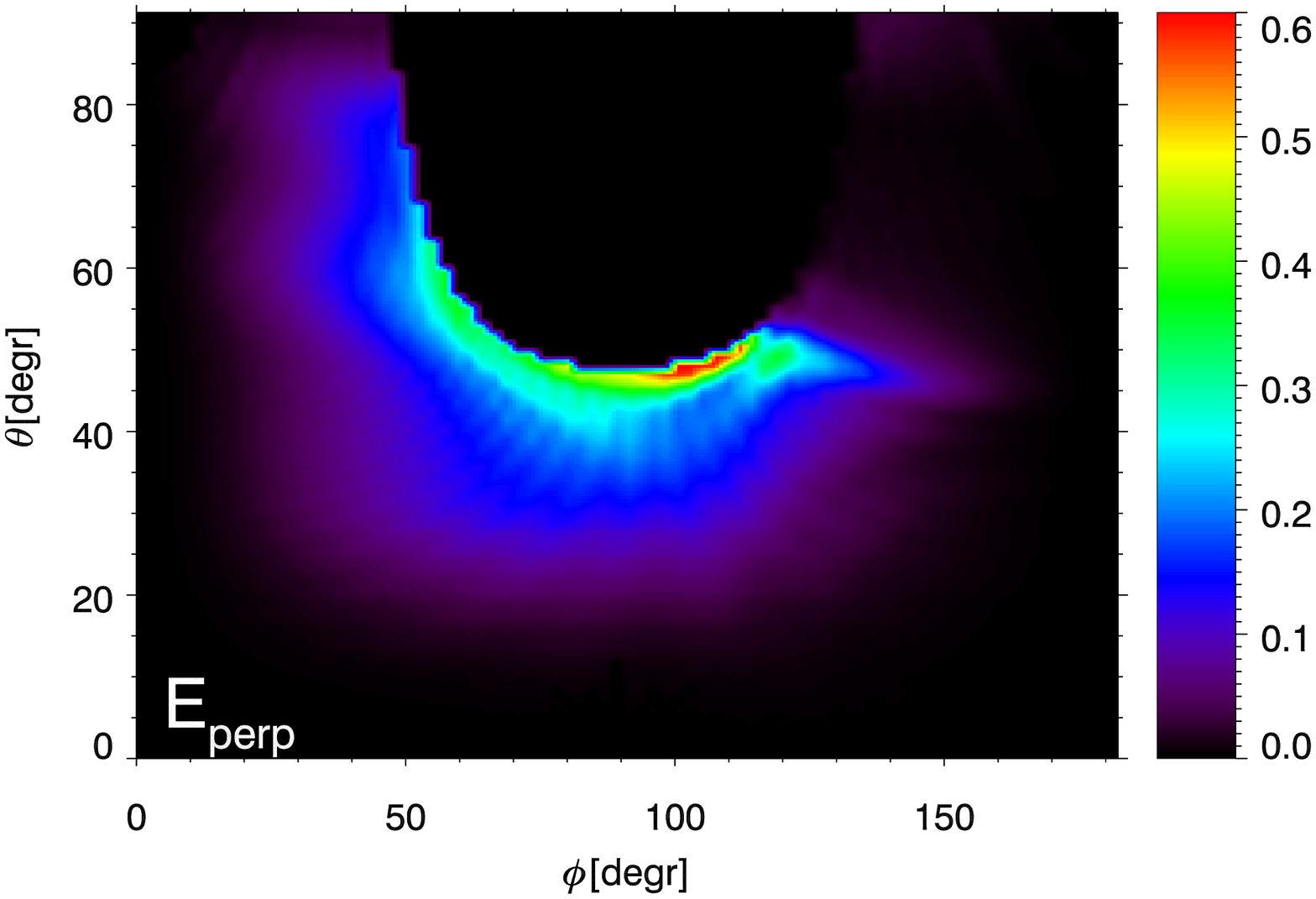}}
\caption[width=\textwidth]{Wave energies at the top of the atmosphere (averages
at heights from 0.4 Mm above the $c_S = v_A$ layer up to z = 1.9 Mm) for the three projected velocity components as a function of inclination and azimuth of the sunspot magnetic field lines at the corresponding horizontal locations. Left panel: acoustic flux; middle panel: magnetic flux; right panel: magnetic flux due to Alfv\'en waves. The units of the color coding are $10^6$ erg cm$^{-2}$s$^{-1}$. In the left panel the regions with negative acoustic vertical flux have been masked.}
\label{fig:energy_angulo}
\end{figure*}

The wave energies have been plotted as function of inclination and azimuth of the sunspot magnetic field lines at the corresponding horizontal locations in Figure \ref{fig:energy_angulo}. This format allows a direct comparison with \citet{Cally+Goossens2008} Figure 2, \citet{Khomenko+Cally2011} Figure 4, and \citet{Khomenko+Cally2012} Figure 5. In the left panel the angles corresponding to the positions with negative acoustic flux has also been masked. It shows that the maximum slow wave energy is obtained when the direction of the wave incidence forms an angle $\phi=0^o$ with the magnetic field, for an inclination around $\theta=25^o$. This result is consistent with the 2D analysis from \citet{Schunker+Cally2006}, where they shown that around this field inclination the attack angle, \ie, the angle between the wavevector and the magnetic field at the conversion layer, is small and produce an enhacement of the conversion from the fast acoustic to the slow acoustic wave. The region where the conversion to the slow mode is efficient extends to higher azimuths, although it decreases significantly for $\phi$ higher than $60^o$. The distribution of the slow mode energy as a function of $\theta$ and $\phi$ shows an excellent agreement with Figure 2 from \citet{Cally+Goossens2008}. The only difference appears at higher azimuths, which correspond to the left part of the sunspots, where the reflection of the slow mode produced by the temperature increase complicates the evaluation of the upward propagating slow mode energy in this simulation of conversion in a realistic sunspot model.

The Alfv\'en wave energy increases toward the periphery of the sunspot, except along $y=0$ Mm (right panel of Figure \ref{fig:energy_XY}). Its maximum value is obtained very close to the boundary of the computational domain around $x=0$ Mm. Its representation as a function of inclination and azimuth (right panel of Figure \ref{fig:energy_angulo}) reveals that the conversion to the Alfv\'en mode is efficient at inclinations above $40^o$ and azimuths between $50^o$ and $120^o$. At $\phi\approx50^o$, which corresponds to the right side of the sunspot, the Alfv\'en energy extends to higher inclinations than the left part of the sunspot. This is also visible in Figure \ref{fig:energy_XY}. The distribution of the Alfv\'en energy with $\theta$ and $\phi$ is remarkably similar to the one found previously from the 3D analysis by \citet{Cally+Goossens2008} in homogeneous fields. However, a few differences arise. In \citet{Cally+Goossens2008} the highest conversion to Alfv\'en waves appears at $\theta\approx40^o$, while in our simulations the maximum energy is shifted toward higher inclinations. A similar shift in the position of the maximum was obtained from the 2.5D simulations of \citet{Khomenko+Cally2011} in homogeneous magnetic fields. Since the Alfv\'enic energy peaks so close to the boundary of our simulations, it is hard to say if the maximum conversion occurs at $\theta\approx47^o$, as shown by the plot, or at even higher inclinations. Moreover, the maximum energy appears in the left part of the sunspot, corresponding to $\phi\approx100^o$. This differs from the analysis of homogeneous magnetic fields, where the maximum is located at $\phi=60^o$ \citep{Cally+Goossens2008,Khomenko+Cally2011} or at $\phi=80^o-90^o$ \citep{Cally+Hansen2011}. In the 2.5D simulations in a sunspot model from \citet{Khomenko+Cally2012} the energy of the Alfv\'en wave peaks at azimuths between $70^o$ and $100^o$.

Middle panel of Figure \ref{fig:energy_angulo} shows some fast wave energy at high inclinations.

\section{Discussion and conclusions}
\label{sect:scattering}
 
In this paper we have studied the conversion from fast to Alfv\'en modes in a realistic sunspot-like atmosphere by means of 3D numerical simulations. This study provides a direct comparison with analytical works \citep{Cally+Goossens2008, Cally+Hansen2011} as well as numerical simulations \citep{Khomenko+Cally2011,Khomenko+Cally2012}.

With regards to the fast-to-slow conversion, the  efficiency of the transformation peaks at inclinations around 25$^o$ and low azimuths, and it decreases at higher azimuths. This result shows a good agreement with previous works \citep{Schunker+Cally2006, Cally+Goossens2008} in homogeneous magnetic field configurations. However, the sunspot-like structure of this simulation produce some differences in the wave modes of the upper atmosphere. In the central and left part of the sunspot the simulation shows downward propagating slow waves, which are reflected due to the chromospheric increase of the temperature. A similar result was found in the 2.5D simulations from \citet{Khomenko+Cally2012}, although in our case the downward flux near the axis is enhanced because of the higher increase of the temperature at the center of the sunspot. In \citet{Khomenko+Cally2012} simulations the closest plane to the axis of the sunspot is located at a distance of 7.5 Mm, where the increase of the temperature in the transition region is lower (Figure \ref{fig:temperature}).

The simulation presented in this paper supposes a step forward in the study of the conversion to the Alfv\'en mode, since it accounts for full 3D. The work by \citet{Khomenko+Cally2012} was limited to 2.5D, and only a few vertical slices across the sunspot model at several distances from the axis were analyzed. Since the fast-to-Alfv\'en conversion depends strongly on the angle between the direction of wave propagation and the magnetic field, a full 3D simulation is required to retrieve the complete description. The simulation shows that the highest Alfv\'enic wave energy is obtained at high inclinations and for azimuths between $50^o$ and $120^o$. The efficiency of the conversion obtained for the sunspot model shows a similar pattern to the homogeneous magnetic fields from \citet{Cally+Goossens2008}, although the maximum of the magnetic energy of Alfv\'en waves is shifted toward more inclined fields.

A comparison of the total energy in the upper atmosphere of the upward propagating slow acoustic energy and the corresponding to the Alfv\'en mode reveals that the former one is around two times higher. This result remarks the important role of the Alfv\'en energy at the chromosphere. At $\theta\approx 25^o$ the acoustic energy is much higher than the Alfv\'en energy, especially at low azimuths where its conversion is more efficient. However, at higher inclinations the Alfv\'en energy becomes dominant and at $\theta=47^o$ and $\phi$ around $90^o$ it is more than 7 times higher than the acoustic energy. The horizontal size of the computational domain prevent us to measure the energy at higher inclinations for these azimuths, but from the tendency shown in the computed region one would expect a strongest dominance of the Alfv\'en energy at higher inclinations. Moreover, the vertical limitation of the domain also reduces the energy in the Alfv\'en mode. At the 5 mHz frequency of this simulation the fast-to-Alfv\'en conversion region spans over 20 scale heights \citep{Cally+Hansen2011}. Since our computational box has a much lower size, only a small fraction of the conversion is completed in our domain. The same limitation affects the simulations from \citet{Khomenko+Cally2012}. However, in our case we have found a higher energy in the Alfv\'en wave due to the higher inclinations covered by our computational domain. 

Despite of the dependence of the conversion with the azimuth, due to the stochastic direction of the wave propagation in the Sun, one would expect to find an efficient conversion to the Alfv\'en wave at all the locations surrounding a sunspot with a certain inclination. Near the umbra-penumbra boundary, the waves whose attack angle is below $90^o$ will generate upward propagating Alfv\'en waves, while the incident waves which propagate in the opposite direction will produce downward Alfv\'en waves. In this way, in these regions upgoing and downgoing waves will coexist. On the other hand, at the higher inclinations of the penumbra, the energy flux of the Alfv\'en mode is only composed by upward propagating waves for all the directions of incidence of the fast acoustic wave where the conversion to the Alfv\'en wave is efficient, as shown by the right panel from Figure \ref{fig:fluxXY} at $\theta\approx 45^o$. It differs from the situation at $\theta\approx 30^o$, where at some regions of the sunspot there is negative flux. In the case of solar observations, where the incident waves propagate in all directions, the Alfv\'en waves in an annular region surrounding the sunspot with higher inclinations (around $\theta\approx 45^o$) will only consist on upward propagating waves.

The small time step imposed by the high chromospheric Alfv\'en speed has been an important concern in the development of the numerical simulations presented in this paper. Some considerations have been taken into account in the numerical configuration regarding this issue. Firstly, a magnetic field strength of 900 G was adopted at the axis of the sunspot at the photosphere. This value is well below what might be expected in a mature sunspot. A more realistic field strength would lower the height of the $c_S=v_A$ and the fast mode reflection layers. This would allow the fast-to-Alfv\'en conversion to be produced in a bigger region of the computational domain, and might generate an even higher Alfv\'en energy flux at the top of the computational domain. Secondly, this top boundary has been set at the chromosphere. The recent work by \citet{Hansen+Cally2012} has shown the interesting effects of the chromosphere-corona transition region on fast-to-Alfv\'en conversion. They found that the reflection of the Alfv\'en waves at the transition region \citep{Uchida+Sakurai1975} is sensitive to the distance between the fast reflection point and the transition region. The Alfv\'en flux that can reach the corona is increased when this distance is small. In the present work the computational box does not include the transition region, since it would produce an even smaller time step, compromising the realization of this simulation using a reasonable amount of computational time. It would be interesting to extend the simulations to the corona and adopt a realistic magnetic field strength, but it is out of the scope of this paper and shall be accounted for in forthcoming works.

\acknowledgements  I would like to thank Elena Khomenko, Paul Cally, Manuel Collados, and Ashley Crouch for their examination and suggestions to an early version of this paper. This work used LaPalma supercomputer at Centro de Astrof\'{\i}sica de La Palma and MareNostrum supercomputer at Barcelona Supercomputing Center (the nodes of Spanish National Supercomputing Center). This work is supported by NASA contract NNH09CE43C. 

\aareferences

\end{document}